\documentclass[showpacs,preprintnumbers,amsmath,amssymb,twocolumn,superscriptaddress,prb]{revtex4}
\usepackage{graphicx}
\usepackage{amsfonts}
\usepackage{amsmath, amsthm, amssymb}
\usepackage{dsfont}
\usepackage{times}
\usepackage{subfigure}

\newcommand{\vecr}{\boldsymbol{r}}

\newcommand{\e}[1]{\mathrm{e}^{#1}}

\newcommand{\bq}{\begin{equation}}
\newcommand{\eq}{\end{equation}}

\newcommand{\g}{\underline{\gamma}}
\newcommand{\gt}{\underline{\tilde{\gamma}}}
\newcommand{\N}{\underline{\mathcal{N}}}
\newcommand{\Nt}{\underline{\tilde{\mathcal{N}}}}

\newcommand{\ie}{\textit{i.e. }}
\newcommand{\eg}{\textit{e.g. }}
\newcommand{\etal}{\emph{et al.}}
\def\i{\mathrm{i}}

\begin{document}
\title[Theory of superconducting and magnetic proximity effect in S$\mid$F structures with inhomogeneous magnetization textures and spin-active interfaces]{Theory of superconducting and magnetic proximity effect in S$\mid$F structures with inhomogeneous magnetization textures and spin-active interfaces}

\author{Jacob Linder}
\affiliation{Department of Physics, Norwegian University of
Science and Technology, N-7491 Trondheim, Norway}
\author{Takehito Yokoyama}
\affiliation{Department of Applied Physics, Nagoya University, Nagoya, 464-8603, Japan}
\author{Asle Sudb{\o}}
\affiliation{Department of Physics, Norwegian University of
Science and Technology, N-7491 Trondheim, Norway}

\date{Received \today}
\begin{abstract}
\noindent We present a study of the proximity effect and the inverse proximity effect in a superconductor$\mid$ferromagnet bilayer, taking into account several important factors which mostly have been ignored in the literature so far. These include spin-dependent interfacial phase shifts (spin-DIPS) and inhomogeneous textures of the magnetization in the ferromagnetic layer, both of which are expected to be present in real experimental samples. Our approach is numerical, allowing us to access the full proximity effect regime. In Part I of this work, we study the superconducting proximity effect and the resulting local density of states in an inhomogeneous ferromagnet with a non-trivial magnetic texture. 
 Our two main results in Part I are a study of how Bloch and N\'eel domain walls affect the proximity-induced superconducting correlations and a study of the superconducting proximity effect in a conical ferromagnet. The latter topic should be relevant for the ferromagnet Ho, which was recently used in an experiment to demonstrate the possibility to generate and sustain long-range triplet superconducting correlations. In Part II of this work, we investigate the inverse proximity effect with emphasis on the induced magnetization in the superconducting region as a result of the "leakage" from the ferromagnetic region. It is shown that the presence of spin-DIPS modify conclusions obtained previously in the literature with regard to the induced magnetization in the superconducting region. In particular, we find that the spin-DIPS can trigger an anti-screening effect of the magnetization, leading to an induced magnetization in the superconducting region with \textit{the same sign} as in the proximity ferromagnet.
  \end{abstract}

\maketitle

\section{Introduction}\label{sec:introduction}
The interplay between ferromagnetism and superconductivity has over the past decade attracted much interest from the condensed-matter physics community. Research on superconductor$\mid$ferromagnet (S$\mid$F) heterostructures continues to benefit from great interest, which is fueled by the exciting phenomena arising from a fundamental physics point of view in addition to the prospect of harvesting functional devices in low-temperature nanotechnology.  
\par
There is currently intense activity in this particular research area  (see \eg  Refs.~\onlinecite{bergeretrmp,buzdinrmp} and references 
therein). The interest in S$\mid$F hybrid structures was boosted at the beginning of this millenium, 
primarily due to the theoretical proposition of proximity-induced odd-frequency correlations \cite{bergeret_prl_01} and the experimental 
observation of 0-$\pi$ oscillations in S$\mid$F$\mid$S Josephson junctions. \cite{ryazanov_prl_01} A large amount of work has been 
devoted to odd-frequency pairing (see \eg \cite{volkov_prl_03,bergeret_prb_03,eschrig_prl_03,Braude,asano_prl_07_1,Keizer,fominov_prb_07,yokoyama_prb_07,  asano_prl_07_2,halterman_prl_07,Tanaka,eschrig_jlow_07,linder_prb_08,eschrig_nphys_08,halterman_prb_08,linder_prb_08_2, yada_arxiv_08}) 
and the physics of 0-$\pi$ oscillations (see \eg \cite{bulaevskii_jetp_77,buzdin_pisma_82,koshina_prb_01,kontos_prl_02,buzdin_prb_03,houzet_prb_05,cottet_prb_05,robinson_prl_06,zareyan_prb_06, yokoyamajos_prb_07, houzet_prb_07, crouzy_prb_07,linder_prl_08,champel_prl_08, brydon_prb_08, sperstad_prb_08,volkov_prb_08}) in S$\mid$F heterostructures. The concept of odd-frequency pairing dates back to Refs. \cite{berezinskii_jetp_74,balatsky_prb_92,coleman_prb_93,abrahams_prb_95} and was recently re-examined in Ref. \cite{solenov_arxiv_08}.
\par
So far, the proximity effect has received much more attention than the inverse proximity effect.
In S$\mid$F bilayers, the proximity effect causes superconducting correlations to penetrate into the ferromagnetic region \cite{bergeretrmp}. Similarly, the inverse proximity effect induces ferromagnetic correlations in the superconducting region near the interface region.\cite{Gu,Sillanpaa,Bergeret,Morten} Often, the bulk solution is employed in the superconducting region, such that both the induced magnetic correlations and the self-consistency of the superconducting order parameter are neglected. However, it was shown in Ref. \cite{bergeret_prb_04} that the induction of an odd-frequency triplet component would lead to a finite magnetization in the superconducting region close to the S$\mid$F interface. Prior to this finding, some experimental groups had reported findings which pointed towards precisely such a phenomenon \cite{muehge_physicac_98, garifullin_appl_02}. Very recently, Xia \etal \cite{xia_arxiv_08} presented an experimental observation of the inverse proximity effect in Al/(Co-Pd) and Pd/Ni bilayers by measuring the magneto-optical Kerr effect. Their data could be roughly fitted to the predictions of Ref. \cite{bergeret_prb_04}, and other experiments \cite{Gu,Sillanpaa,muehge_physicac_98, garifullin_appl_02, salikhov_arxiv_08} have also addressed aspects of the inverse proximity effect S$\mid$F bilayers.
\par
In Ref. \cite{kharitonov_prb_06}, the authors investigated the proximity-induced magnetization in the superconducting region of a S$\mid$F bilayer, and found that the magnetization would oscillate in the clean limit (see also Ref. \cite{halterman_prb_04}) and decay monotonously in the diffusive limit, with a sign opposite to the magnetization in the bulk of the ferromagnet. The reason for this screening behavior in the superconductor was attributed to a scenario in which the spin-$\uparrow$ electron of a Cooper pair near the interface would prefer to be located in the ferromagnetic region, while its spin-$\downarrow$ partner would remain in the superconducting region, thus creating a magnetization with an opposite sign compared to the ferromagnet. By considering the weak proximity effect regime in the diffusive limit, both Ref. \cite{bergeret_prb_04} and Ref. \cite{kharitonov_prb_06} arrived at this conclusion. However, it would be desirable to go beyond the approximation of a weak proximity effect employed in previous work to investigate if this may alter how the induced magnetization in the superconducting region behaves. 
\par
Moreover, none of the above works on the inverse proximity effect have properly included an important property which is intrinsic to S$\mid$F interfaces, namely the spin-dependent interfacial phase shifts (spin-DIPS) that occur at the interface. The spin-DIPS have been shown to exert an important influence on various experimentally observable quantities in S$\mid$F bilayers \cite{cottet_prb_05,huertashernando_prl_02,cottet_prb_07}, and should be taken into account. For instance, the anomalous double peak structure in the local density of states (LDOS) in a diffusive S$\mid$F bilayer reported very recently by SanGiorgio \etal  in Ref. \cite{sangiorgio_prl_08} was reproduced theoretically in Ref. \cite{cottet_arxiv_08} by using a numerical solution of the Usadel equation when including the effect of the spin-DIPS.
\par
So far, due to the complexity of the problem, several assumptions have been usually made when treating S$\mid$F hybrid structures.
For instance, 
since the quasiclassical equations become quite complicated for inhomogeneous ferromagnets, they have been 
linearized in most of the previous works. However, presently, the direction of this research field tends 
towards a more realistic description of S$\mid$F structures than the simplified models that mostly have been 
employed up to now. It is obvious that this is a necessary step in order to reconcile the theoretical 
predictions with experimentally observed data. 
\par
Our motivation for this work is to examine the effect of inhomogeneous magnetization textures and spin-DIPS on both the proximity effect and the inverse proximity effect in S$\mid$F bilayers. This is directly relevant to two recent experimental studies \cite{sosnin_prl_06, xia_arxiv_08} which studied the superconducting proximity effect in the conical ferromagnet Ho and the inverse proximity effect in the superconducting region of a S$\mid$F bilayer, respectively. As we shall show in this work, non-trivial magnetization textures and spin-DIPS have profound influence on the physical properties of S$\mid$F bilayers, suggesting that their role must be taken seriously.
\par
We divide this work into two parts which are devoted to the proximity effect in the ferromagnetic region (Part I) and the inverse proximity effect in the superconducting region (Part II). In Part I, we present results where we treat the role of magnetic properties at the interface and the possibility of inhomogeneous magnetization 
thoroughly. 
We study the proximity-induced density of states (DOS) in a S$\mid$F bilayer which takes into account the presence of 
spin-DIPS at the interface and also the possibility of having a non-trivial magnetization texture (such as a domain wall) in 
the ferromagnetic region. In order to access the full proximity effect regime, we do not restrict ourselves to any limiting 
cases. Rather, we employ a full numerical solution of the DOS by means of the quasiclassical theory of superconductivity. We apply 
our theory to two cases of ferromagnets with an inhomogeneous magnetic texture, namely on one hand ferromagnets with domain 
walls and on the other hand conical ferromagnets. 
\par
In Part II, we study numerically and self-consistently the inverse proximity effect in a S$\mid$F bilayer of finite size upon taking properly into account the spin-DIPS that occur at the S$\mid$F interface. Our main objective is to study the influence exerted on the inverse proximity effect by the spin-DIPS. Surprisingly, we find that the spin-DIPS may invert the sign of the proximity-induced magnetization in the superconducting layer compared to the predictions of Refs. \cite{bergeret_prb_04, kharitonov_prb_06}. Consequently, the spin-DIPS can trigger an anti-screening effect of the magnetization, which suggests that their role must be taken seriously in any attempt to construct a theory for the inverse proximity effect in S$\mid$F bilayers. We also explain the basic mechanism behind the sign-inversion induced by the spin-DIPS.
\par
This paper is organized as follows. 
In Section \ref{sec:theoryI}, we present the theoretical framework we use to perform our computations in Part I, namely the quasiclassical theory of superconductivity in the diffusive limit for an inhomogeneous ferromagnet using the Ricatti parametrization. In Section \ref{sec:resultsI},
we present our numerical results for proximity-effect and the local density of states for the two cases of ferromagnets with
domain walls and with conical magnetic textures.  In Section \ref{sec:discussionI}, we present a discussion of our
results obtained in Part I. Moving on to Part II of this work, we introduce a slightly different notation and parametrization for the Green's function in Sec. \ref{sec:theoryII}, which is easier to implement for a homogeneous S$\mid$F bilayer. In Sec. \ref{sec:resultsII}, we present our results for the inverse proximity effect, manifested through an induced magnetization in the superconducting region and in particular how it is influenced by the presence of spin-DIPS. The results for Part II are discussed in Sec. \ref{sec:discussionII}, and we conclude with final remarks in Sec. \ref{sec:summary}. Throughout the paper, we will use boldface 
notation for 3-vectors, $\hat{\ldots}$ for $4\times4$ matrices, and $\underline{\ldots}$ for 
$2\times2$ matrices.

\section{ Proximity effect in a S$\mid$F bilayer with an inhomogeneous magnetization texture}\label{sec:PartI}

\subsection{Theoretical framework}\label{sec:theoryI}

In the first part of our work, we shall consider the proximity effect in the ferromagnetic region of an S$\mid$F bilayer when the magnetization texture is inhomogeneous. This is the case \eg in the presence of a domain-wall structure or conical ferromagnetism, which both will be treated below. We will use the quasiclassical theory of superconductivity \cite{serene}, and consider the diffusive limit described by the Usadel equation \cite{usadel}. 

\subsubsection{Quasiclassical theory and Green's functions}
To account for an inhomogeneous magnetization in the ferromagnet, it is convenient to parametrize 
the Green's function to obtain a simpler set of equations to solve. One possibility is to use a generalized $\theta$-parametrization \cite{ivanov_prb_06}, as follows
\begin{align}
\hat{g} &= \begin{pmatrix}
M_0c\underline{\sigma_0} + (\boldsymbol{M}\cdot\underline{\boldsymbol{\sigma}})s & \underline{\rho}^+ \notag\\
\underline{\rho}^- & -M_0c\underline{\sigma_0} - (\boldsymbol{M}\cdot\underline{\boldsymbol{\sigma}})^*s
\end{pmatrix},\notag\\
\underline{\rho}^\pm &= c[\i(M_z\underline{\sigma_2}-M_y\underline{\sigma_3})\pm M_x\underline{\sigma_0}] \pm M_0\underline{\sigma_1}s,
\end{align}
where $\underline{\sigma_j}$ are the identity $(j=0)$ and Pauli $(j=1,2,3)$ matrices, and \begin{align}
\underline{\boldsymbol{\sigma}} = (\underline{\sigma_1}, \underline{\sigma_2}, \underline{\sigma_3}).
\end{align}
Also, $s\equiv \sinh(\theta)$ and $c\equiv \cosh(\theta)$. The Green's function is then completely determined by the complex functions $\theta$, $M_0$, and $\boldsymbol{M}$ with the additional constraint $M_0^2 -\boldsymbol{M}^2=1$ in order to satisfy $\hat{g}^2=\hat{1}$. 
\par
However, for our purpose we find it both more convenient and elegant to use a Ricatti-parametrization of the Green's function as follows \cite{schopohl_prb_95,konstandin_prb_05}, as follows
\begin{align}\label{eq:g}
\hat{g} &= \begin{pmatrix}
\N(\underline{1}-\g\gt) & 2\N\g \\
2\Nt\gt & \Nt(-\underline{1} + \gt\g) \\
\end{pmatrix}.
\end{align}
This parametrization  facilitates the numerical computations, and also ensures that $\hat{g}^2=\hat{1}$. 
The unknown functions $\g$ and $\gt\g$ are key elements in this parametrization of the Green's function, and  
will be solved for below. Here, $\underline{\ldots}$ denotes a $2\times2$ matrix and 
\begin{align}
\N=(1+\g\gt)^{-1}\; \Nt = (1+\gt\g)^{-1}.
\end{align}
\par
In order to calculate the Green's function $\hat{g}$, we need to solve the Usadel equation with appropriate boundary conditions at $x=0$ and $x=d_F$. The two natural length scales associated with each of the long-range orders are the superconducting and ferromagnetic coherence lengths 
\begin{align}
\xi_S = \sqrt{D_S/\Delta_0},\; \xi_F = \sqrt{D_F/h_0},
\end{align}
 where $\Delta_0$ and $h_0$ denote the bulk values of the gap and the exchange field. We set $D_F=D_S=D$ for simplicity. The Usadel equation reads
\begin{align}\label{eq:usadel}
D\partial(\hat{g}\partial\hat{g}) + \i[\varepsilon\hat{\rho}_3 + \text{diag}[\boldsymbol{h}\cdot\underline{\boldsymbol{\sigma}},(\boldsymbol{h}\cdot\underline{\boldsymbol{\sigma}})^\mathcal{T}], \hat{g}]=0,
\end{align}
and is supplemented with the boundary conditions \cite{cottet_prb_05,huertashernando_prl_02}
\begin{align}
2\zeta d_F\hat{g} \partial \hat{g} = [\hat{g}_\text{BCS}, \hat{g}] + \i (G_\phi/G_T) [\text{diag}(\underline{\tau_3}, \underline{\tau_3}), \hat{g}]
\end{align} 
at $x=0$ where the interface is spin polarized along the z-axis, and $ \hat{g}\partial\hat{g}=\hat{0}$ at $x=d_F$. Here, $\partial \equiv \frac{\partial}{\partial x}$ and we define
\begin{align}
\zeta=R_B/R_F
\end{align}
as the ratio between the resistance of the barrier region and the resistance in the ferromagnetic film (note that $R_B = G_T^{-1}$). The barrier conductance is given by \cite{cottet_prb_05} 
\begin{align}
G_T = G_Q\sum_n^N T_n,
\end{align}
 where $G_Q = e/h$ and $T_n$ is the transmission coefficient for channel $n$. The boundary conditions Eqs. (\ref{eq:bcF}) and (\ref{eq:bcS}) are derived under the assumption that $T_n \ll 1$, but this does not necessarily mean that the barrier conductance is small since there may be a large total number of channels $N$ through which transport may take place. The parameter $G_\phi$ describes the spin-DIPS taking place at the F side of the interface.\cite{Brataas} Since its exact value depend on the microscopic properties of the barrier region, they are here treated phenomenologically. We finally underline that the boundary conditions above are valid for planar diffusive contacts.

\par
Since we employ a numerical solution, we have access to study the full proximity effect regime and also an, in principle, 
arbitrary spatial modulation $h=h(x)$ of the exchange field. This is desirable in order to clarify effects associated 
with non-uniform ferromagnets, such as spiral magnetic ordering or the presence of domain walls. Inserting 
Eq. (\ref{eq:g}) into Eq. (\ref{eq:usadel}), we obtain the transport
equation for the unknown function $\g$ (and hence $\gt\g$)
\begin{align}
D[\partial^2\g +(\partial\g)\underline{\tilde{\mathcal{F}}}(\partial\g)] + \i[2\varepsilon\g + \boldsymbol{h}\cdot(\underline{\boldsymbol{\sigma}}\g - \g\underline{\boldsymbol{\sigma}}^*)] = 0,
\end{align}
with $\underline{\tilde{\mathcal{F}}} = -2\Nt\gt$. The boundary condition at $x=0$ reads
\begin{align}
2\zeta d_F\partial_x\g &= [2c\g -s\i\underline{\tau_2} + \g(s\i\underline{\tau_2})\g] \notag\\
&+ \i (G_\phi/G_T)(\underline{\tau_3}\g - \g\underline{\tau_3}),
\end{align}
while $\partial_x\g = 0$ at $x=d$. For $\gt$, we obtain
\begin{align}
D[\partial^2\gt +(\partial\gt)\underline{\mathcal{F}}(\partial\gt)] + \i[2\varepsilon\gt + \boldsymbol{h}\cdot(\gt\underline{\boldsymbol{\sigma}} - \underline{\boldsymbol{\sigma}}^*\gt)] = 0,
\end{align}
with the corresponding boundary condition
\begin{align}
2\zeta d_F\partial_x\gt &= [2c\gt -s\i\underline{\tau_2} + \gt(s\i\underline{\tau_2})\gt] \notag\\
&- \i (G_\phi/G_T)(\underline{\tau_3}\gt - \gt\underline{\tau_3}).
\end{align}
We have defined $\underline{\mathcal{F}} = -2\N\g$. Note that we use the bulk solution in the superconducting region, which is a good approximation when assuming that the superconducting region is much less disordered than the ferromagnet and when the interface transparency is small, as considered here (see detailed discussion in Sec. \ref{sec:discussionI}). One finds that 
\begin{align}
\g_\text{BCS}=\gt_\text{BCS} = \begin{pmatrix}
0 & s/(1+c)\\
-s/(1+c) & 0 \\
\end{pmatrix}.
\end{align}
The normalized DOS is finally evaluated by 
\begin{align}
N(\varepsilon)/N_0 = \text{Tr}\{\text{Re}[\N(1-\g\gt)]\}/2.
\end{align}
In what follows, we will omit the effect of spin-flip and spin-orbit scattering to reduce the number of parameters in 
the problem. In comparison with real experimental data, however, the effects of these pair-breaking mechanisms are 
easily included in our framework by adding two terms $\hat{\sigma}_\text{sf}$ and $\hat{\sigma}_\text{so}$ in 
Eq. (\ref{eq:usadel}) (see \eg Ref. \cite{linder_prb_08_2} for a detailed treatment). In this paper, we will 
focus on the role of the phase-shifts obtained at the interface due to the spin-split bands and the 
inhomogeneity of the exchange field in the ferromagnet.

\begin{figure}[h!]
\centering
\resizebox{0.49\textwidth}{!}{
\includegraphics{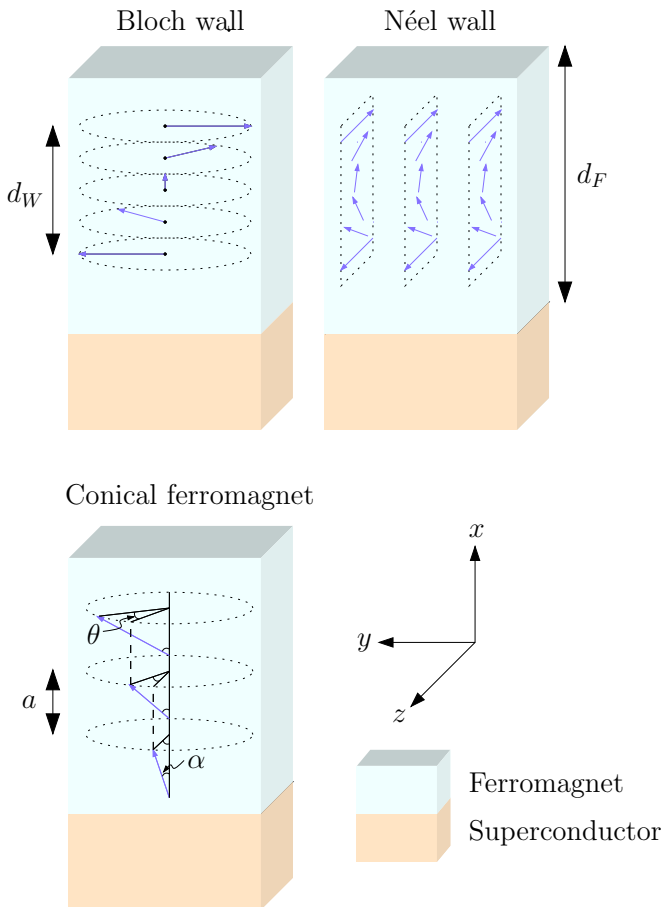}}
\caption{(Color online) The three types of inhomogeneous ferromagnets we will consider in this work: Bloch walls, N\'eel walls, and conical ferromagnets (such as Ho).}
\label{fig:model}
\end{figure}

\subsubsection{Inhomogeneous magnetization}
We will consider three types of inhomogeneous magnetic structures: Bloch walls, N\'eel walls, and conical ferromagnets 
(see Fig. \ref{fig:model}). An example of the latter is the rare-earth heavy fermion elemental magnet Ho, although 
we hasten to add that while Ho features strong ferromagnet, we will consider the weakly ferromanetic case. These 
structures are shown in Fig. \ref{fig:model} and 
are to be contrasted with the usual assumption of a homogeneous exchange field in the ferromagnetic region. For
 the first two cases, the domain wall has a width $d_W$ and is taken to be located at the center of the 
 ferromagnetic region $(x=d_F/2)$. The Bloch wall is thus modelled by 
\begin{align}
\mathbf{h} = h(\cos\theta\hat{\mathbf{y}}+\sin\theta\hat{\mathbf{z}}),
\end{align}
while $\hat{\mathbf{y}}\to \hat{\mathbf{x}}$ for the N\'eel wall. Here, we have defined 
\begin{align}
\theta=-\arctan[(x-d_F/2)/d_W],
\end{align}
similarly to Ref. \cite{konstandin_prb_05}.
\par
In the case of a conical ferromagnet,
cf. Fig. \ref{fig:model}, the magnetic moment belongs to a cone. In Ho, the opening angle is
$\alpha=4\pi/9$ and the magnetic moment then rotates like a helix along the $c$-axis with a
turning angle $\theta=\pi/6$ per interatomic layer with distance $a$ (see Ref. \cite{sosnin_prl_06} for a further discussion). 
Above 21 K, the conical
ferromagnetic structure transforms into a spiral antiferromagnetic
structure. Instead of using an abrupt change in the magnetization direction at each interatomic layer, we will model 
this transition smoothly since the effective field felt between the layers probably should be a weighed superposition 
of the exchange fields from the two closest layers. In the ferromagnetic phase, the spatial variation of the exchange 
field may thus be written as
\begin{align}
\mathbf{h} = h\Big[&\cos\alpha\hat{\mathbf{x}} + \sin\alpha\Big\{\sin\Big(\frac{\theta x}{a}\Big)\hat{\mathbf{y}} +\cos\Big(\frac{\theta x}{a}\Big)\hat{\mathbf{z}}\Big\}\Big].
\end{align}

\subsection{Results}\label{sec:resultsI}
In what follows, we will choose the parameters of our model, corresponding to a realistic experimental setup in order to 
make our study directly relevant for experiments on S$\mid$F bilayers. The numerical treatment makes use of built-in routines in MATLAB for a two-point boundary value problem for an ordinary differential equation. More specifically, we use a finite difference code which implements a three-stage Lobatto-Illa formula. An initial guess for the Ricatti-matrices is supplied with fixed boundary conditions, and the Usadel equation is then solved in the entire ferromagnetic region.
\par
In the first part of this section, we will study the effect of domain walls in weak ferromagnets. Weak ferromagnetic alloys 
such as PdNi or CuNi are commonly employed in experiments, and the corresponding exchange field $h$ depends on the 
concentration of Ni, reaching up to tens of meV. 
The modification of the DOS is most dramatic in the case when the energy scales for the superconductivity and the 
ferromagnetism are of the same order, $h\sim\Delta$. This scenario appears to have been realized in Ref.~\onlinecite{ryazanov_prl_01} where Cu$_{1-x}$Ni$_x$ with $x=0.44$ was used. The diffusion constant in the weakly ferromagnetic alloys is usually of order $D \sim 10^{-4}$ m$^2$/s. The superconducting region is considered to act as a reservoir with thickness $d_S\gg\xi_S$, while we fix the thickness of the ferromagnetic region at $d_F/\xi_S = 0.5$. This typically corresponds to a thickness of the ferromagnetic layer $\sim$ 10 nm. The remaining parameters are then the domain wall thickness $d_W$ and the term $G_\phi$ accounting for the spin-dependent phase-shifts at the interface. Below, we will contrast a thin domain wall $(d_W\ll d_F)$ with a thick domain wall $(d_W \simeq d_F)$ and investigate the role of $G_\phi$. In what follows, we choose $\zeta=5$ corresponding to a situation where $R_B\gg R_F$.
\par
In the second part of this section, we will study conical ferromagnetism, of a similar kind to that realized in the heavy rare earth element 
Holmium (Ho) under certain conditions. Recently, it was strongly suggested by experimental data that a long-range triplet superconducting
component was generated and sustained in a superconductor$\mid$Ho proximity structure \cite{sosnin_prl_06}. The experimental 
samples used in Ref. \cite{sosnin_prl_06} did not appear to fall into the 
diffusive motion regime, since Ho is a strong ferromagnet. More specifically, it was estimated that $h\tau \simeq 10$ in Ref. \cite{sosnin_prl_06}, suggesting that one would have to revert to the more general Eilenberger equation in order to study the proximity effect in Ho. In 
this work, we will study a conical ferromagnet under the assumption that the diffusive limit is reached. For the actual structure of 
the magnetization, we choose the same parameters for Ho as those reported in Ref. \cite{sosnin_prl_06}: $\alpha=4\pi/9$, $\theta=\pi/6$, 
and $a=0.526$ nm (see Fig. \ref{fig:model}). However, we  choose the exchange field much weaker than in Ho, in order to justify the 
Usadel approach. Thus, our results may not be directly applicable to Ho.  While in Ref. \cite{sosnin_prl_06} it was estimated that 
$h \sim 1$ eV, corresponding to an exchange field comparable in magnitude with the Fermi energy, we choose $h/\Delta_0=5$ in 
our study of conical ferromagnetism to ensure the validity of the quasiclassical approach. Assuming that $\xi_S = 20$ nm, 
which should be reasonable for a moderately disordered conventional superconductor, we obtain $a/\xi_S = 0.0263$.

\subsubsection{Domain wall}

Before proceeding to a dissemination of our results, it should be noted that we find identical results for the Bloch and N\'eel wall cases. 
This seems reasonable, since the only difference between those two cases is that the $y$-component of the magnetization is exchanged with 
the $x$-component. The long-range triplet component comes about as long as only one of these is non-zero, and it does not matter which 
one it is. It is also necessary for the magnetization to vary directionally with the $x$-coordinate in order to generate the inhomogeneity 
required for the long-range triplet component. Note that the $z$-component of the magnetization is the same for the Bloch and N\'eel 
walls. In what follows, we only consider the Bloch wall configuration since the results for the N\'eel wall are identical. We also 
note that in our study, the magnetization is always inhomogeneous in the direction perpendicular to the interface, i.e. upon 
penetrating into the ferromagnetic region. In the case where the inhomogeneity of the magnetization is in the transverse direction 
(parallell to the interface), i.e. there is no variation in the $x$-direction, the proximity effect does not become long-ranged 
even if equal-spin correlations may be generated \cite{champel_prl_08}. The general condition for a long-range proximity effect 
is that there exists a misalignment between the triplet anomalous Green's function vector and the exchange field.
\par
We first study the thin-domain wall case $d_W/d_F=0.2$. To begin with, we shall consider the energy-resolved DOS in the center of the domain wall $(x=d_F/2)$ for several values of the exchange field. This is shown in Fig. \ref{fig:thinE}. As seen, the zero-energy DOS is enhanced in all cases due to the presence of odd-frequency correlations.\cite{asano_prl_07_1,Braude,yokoyama_prb_07, yokoyama_prb_05} The influence of the spin-DIPS $(G_\phi)$ seems to be an induction of additional peak features in the subgap regime. This effect is most pronounced at low exchange fields (in particular $h/\Delta_0=0.5$ in Fig. \ref{fig:thinE}). A possible physical explanation for the additional peak features in the LDOS may be the fact that $G_\phi$ acts as an effective exchange field in both the superconducting and ferromagnetic layers.\cite{huertashernando_prl_02} It thus conspires with the intrinsically existing exchange field in the ferromagnetic layer to yield a modified value of the total exchange field. This explanation is consistent with the fact that the position of the peaks change upon increasing $G_\phi$. More specifically, the spin-DIPS appear to enhance the exchange field since the peaks move outwards toward the gap edge.
\par
Next, we investigate the thick domain wall case, and choose $d_W/d_F=0.8$. In Fig. \ref{fig:thickE}, we again consider the energy-resolved LDOS in the middle of the ferromagnetic layer $(x/d_F=0.5)$ for three different values of the exchange field. Upon comparison with Fig. \ref{fig:thinE}, it is seen that the general trend upon increasing the domain wall thickness is an overall enhancement of the proximity effect. The qualitative features in Fig. \ref{fig:thickE} are very similar to those in the thin domain wall case, but the enhancement at zero-energy tends to be larger, particularly so for large values of $h/\Delta_0$. Again, it is seen that the effect of the spin-DIPS is a modification of the total exchange field, amounting to a double-peak structure at subgap energies in the LDOS.

\begin{widetext}
\text{ }\\
\begin{figure}[h!]
\centering
\resizebox{1.01\textwidth}{!}{
\includegraphics{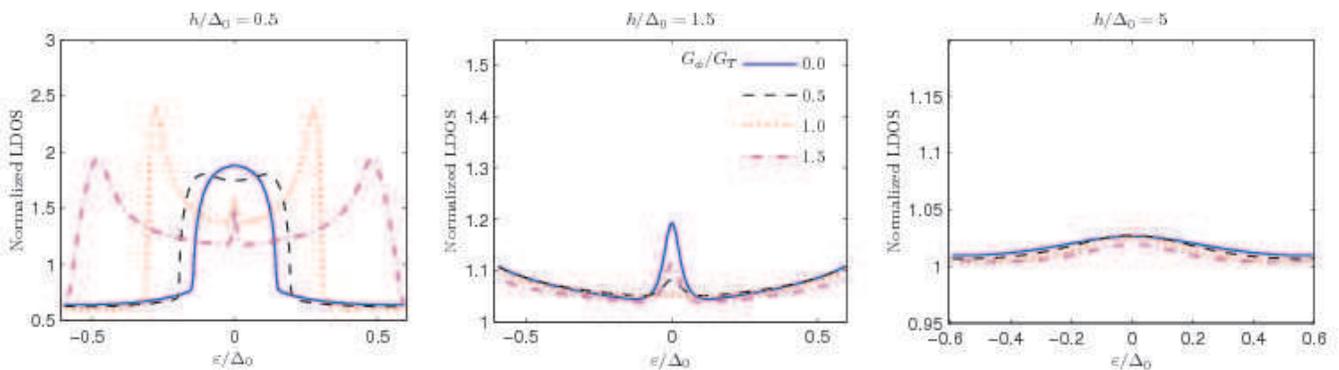}}
\caption{(Color online) Plot of the energy-resolved LDOS evaluated at $x/d_F=0.5$ in the case of a thin domain-wall $d_W/d_F=0.2$. We consider three values of the exchange field $h$ and also investigate how the LDOS changes with the phase-shift $G_\phi$ at the interface.}
\label{fig:thinE}
\end{figure}
\text{ }\\
\begin{figure}[h!]
\centering
\resizebox{1.01\textwidth}{!}{
\includegraphics{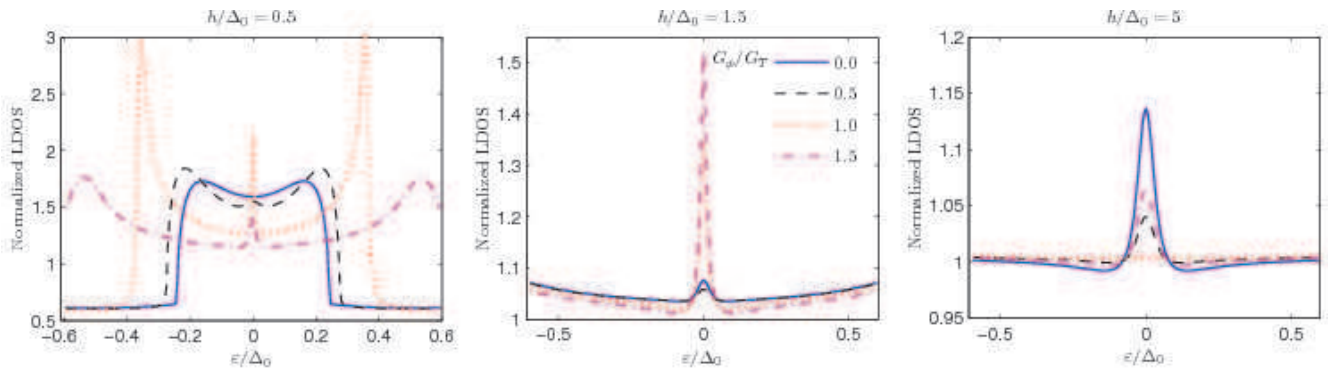}}
\caption{(Color online) Plot of the energy-resolved LDOS evaluated at $x/d_F=0.5$ in the case of a thick domain-wall, $d_W/d_F=0.8$. We consider three values of the exchange field $h$ and also investigate how the LDOS changes with the phase-shift $G_\phi$ at the interface.}
\label{fig:thickE}
\end{figure}
\end{widetext}

It is also interesting to consider the spatial dependence of the zero-energy DOS in the ferromagnetic region. By using local STM-techniques, it is possible to probe the DOS at (in principle) any location in the ferromagnetic film. The specific choice of $\varepsilon=0$ is particularly interesting in terms of the DOS, since it is strongly influenced by the presence of odd-frequency correlations. As pointed out in Refs.~\onlinecite{yokoyama_prb_07, linder_prb_08_2}, the behaviour of the DOS at $\varepsilon=0$ may be interpreted as a competition between spin-singlet even-frequency correlation and spin-triplet odd-frequency correlation. The former tend to give a minigap in the DOS for subgap energies, while the latter yields a zero-energy peak in the DOS. Clearly, these two effects are competing with each other since they have a destructive interplay. In the present case, one would expect that the domain wall structure should favor the generation of the odd-frequency triplet components, thus enhancing the LDOS. This conjecture is supported by Figs. \ref{fig:thinE} and \ref{fig:thickE}. 
\par
\begin{figure}[h!]
\centering
\resizebox{0.47\textwidth}{!}{
\includegraphics{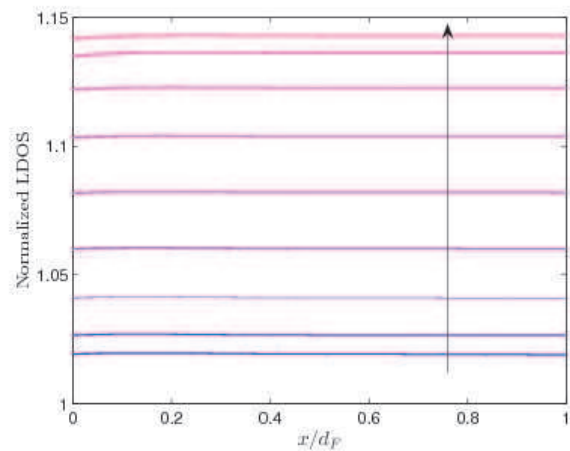}}
\caption{(Color online) Plot of the zero-energy LDOS induced in the ferromagnet for $h/\Delta_0=5$ and $d_F/\xi_S=0.5$. The lines correspond to $d_W/d_F$ in the range $[0.1,0.9]$ in steps of $0.1$ along the arrow. Here, $G_\phi$ is set to zero.}
\label{fig:DOS_x}
\end{figure}

In Fig. \ref{fig:DOS_x}, we plot the spatially-resolved LDOS at $\varepsilon=0$ for several values of $d_W$ to probe directly how the odd-frequency correlations are affected by the domain wall thickness. As compared to Figs. \ref{fig:thinE} and \ref{fig:thickE}, we normalize the LDOS on its value at $x=0$ in Fig. \ref{fig:DOS_x} for easier comparison between different values of $d_W$, and choose $G_\phi=0$. From the plot, it is clear that the thicker the domain wall, the more strongly enhanced the zero-energy DOS. This also supports the notion that the magnetically inhomogeneous structure favors the generation of odd-frequency triplet components. The concomitant enhancement of the DOS may then be seen at increasingly larger penetration depths in the ferromagnet when the domain wall thickness is increased.

\subsubsection{Conical ferromagnetism}
We now turn to a study of how the superconducting proximity effect is manifested in a ferromagnet with a conical magnetization such as Ho. We fix the exchange field at $h/\Delta_0=5$ and study how the DOS changes upon increasing the ferromagnetic layer thickness. The motivation for this is to obtain a better understanding of how the DOS changes when only the long-range triplet components are present in the sample. In an inhomogeneous ferromagnet, the singlet component and the $S_z=0$ triplet component are short-ranged, and penetrate in a distance $\xi_F = \sqrt{D/h}$ into the ferromagnet. The $S_z=\pm1$ triplet components, however, are not subject to the pair-breaking effect originating with the Zeeman splitting, and can thus penetrate a much longer distance $\xi_N=\sqrt{D/T}$ into the ferromagnet, where $T$ is temperature. Therefore, by making the ferromagnetic layer thick enough, one can be certain that there is no contribution from either the singlet or $S_z=0$ triplet components. Since we have  chosen $h/\Delta_0=5$, we find that the penetration depth of these components in the ferromagnetic layer should be $0.44\xi_S$. 
\par
We next turn to a study of the proximity-induced LDOS. In Fig. \ref{fig:conical_E}, we plot the energy-resolved LDOS for three layer thicknesses: \textit{i)} $d/\xi_S=0.1$, \textit{ii)} $d/\xi_S=0.5$, and \textit{iii)} $d/\xi_S=0.9$. In case \textit{i)}, both short-ranged and long-ranged components should contribute significantly to the LDOS. In case \textit{ii)}, the long-ranged components should dominate over the short-ranged ones, while finally in case \textit{iii)} only long-ranged components remain. This is because we evaluate the energy-resolved DOS at the F$\mid$I interface, $x=d_F$, as was also done in the experiment of Refs. \cite{kontos_prl_01,sangiorgio_prl_08}.
\par
As seen in case \textit{ii)} and \textit{iii)}, a pronounced zero energy peak is present, bearing witness of the odd-frequency correlations in the system. The peak is more pronounced with increasing thickness, since the long-range triplet correlations dominate over the even-frequency singlet Green's function as the thickness increases. However, case \textit{i)} is qualitatively different from the two other thicknesses. In this case, the low-energy LDOS is completely suppressed in the regime $G_\phi/G_T<1$, and suddenly reappears for $G_\phi/G_T>1$. It is very interesting to note that the same effect was recently discovered for an S$\mid$N junction with a magnetically active interface \cite{linder_submitted_08}, but in that case the effect was completely independent of the junction thickness.

\begin{widetext}
\text{ }\\
\begin{figure}[h!]
\centering
\resizebox{1.01\textwidth}{!}{
\includegraphics{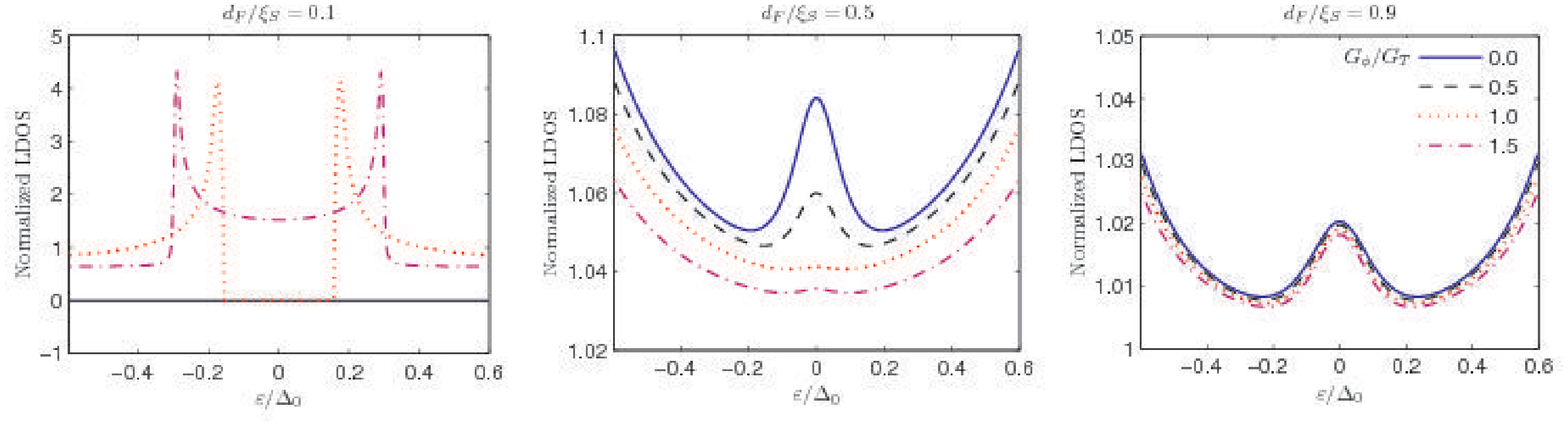}}
\caption{(Color online) Plot of the LDOS at $x=d_F$ for a conical ferromagnet with $h/\Delta_0=5$ for several values of the ferromagnetic layer thickness $d_F$. In each case, we investigate the role of the spin-DIPS $(G_\phi)$ at the S$\mid$F interface. }
\label{fig:conical_E}
\end{figure}
\end{widetext}

\begin{figure}[h!]
\centering
\resizebox{0.45\textwidth}{!}{
\includegraphics{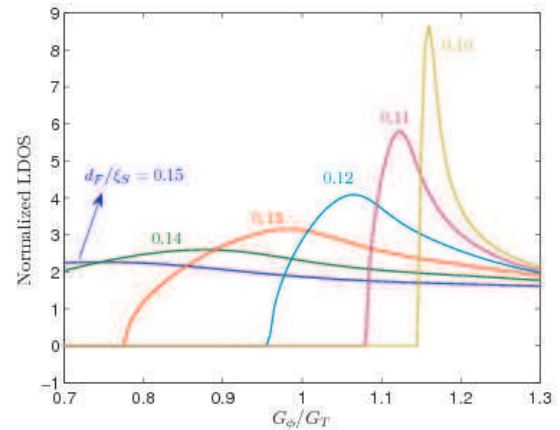}}
\caption{(Color online) Plot of the zero-energy LDOS at $x=d_F$ for a conical ferromagnet with $h/\Delta=5$ as a function of the normalized spin-DIPS parameter $G_\phi/G_T$. Below a critical value for $d_F$, a qualitatively different behavior is observed for the LDOS. }
\label{fig:conical_G2}
\end{figure}

\begin{figure}[h!]
\centering
\resizebox{0.45\textwidth}{!}{
\includegraphics{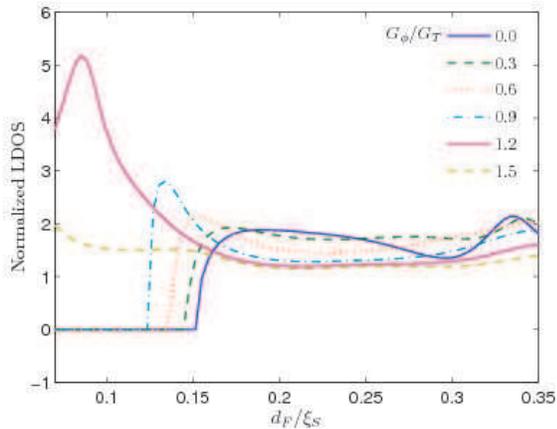}}
\caption{(Color online) Plot of the zero-energy LDOS at $x=d_F$ as a function of $d_F/\xi_S$ for several values of $G_\phi$. As seen, there is a critical thickness at which the zero-energy LDOS becomes non-zero. We have used $h/\Delta_0=5$.}
\label{fig:conical_d}
\end{figure}

\begin{figure}[h!]
\centering
\resizebox{0.48\textwidth}{!}{
\includegraphics{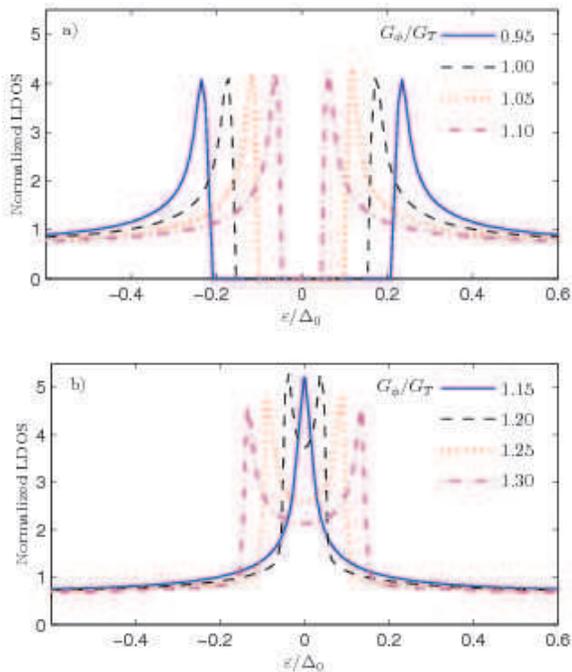}}
\caption{(Color online) Plot of the LDOS at $x=d_F$ for a conical ferromagnet with $h/\Delta_0=5$ and $d_F/\xi_S=0.1$ for several values of the spin-DIPS $(G_\phi)$ at the S$\mid$F interface. In a), $G_\phi/G_T$ is below the critical value, while in b) $G_\phi/G_T$ is larger than the critical value.}
\label{fig:conical_G}
\end{figure}

In order to investigate this effect further, we focus on the zero-energy LDOS in the thin junction case in Fig. \ref{fig:conical_G2}. As seen, for sufficiently thin layers $d_F/\xi_S \ll 1$, an abrupt crossover takes place at a critical value of $G_\phi/G_T$, qualitatively altering the LDOS at zero-energy. Remarkably, we find that a similar transition takes place upon increasing the ferromagnetic layer thickness. Consider a plot of the zero-energy LDOS in Fig. \ref{fig:conical_d} as a function of $d_F/\xi_S$. As seen, at a critical layer thickness, the zero-energy LDOS rises abruptly from zero and acquires the usual oscillating behavior. To see how the full energy-resolved LDOS evolves with increasing $G_\phi$ for a fixed thickness $d_F/\xi_S = 0.1$, consider Fig. \ref{fig:conical_G}. As seen, the LDOS changes qualitatively above a critical value of $G_\phi/G_T \simeq 1.14$.

To summarize the findings of Figs. \ref{fig:conical_G2}, \ref{fig:conical_d}, and \ref{fig:conical_G}, we have found that there is an abrupt crossover from a fully suppressed LDOS to a finite LDOS which appears at a critical thickness of the ferromagnetic layer, and the particular value of the critical thickness depends on the value of $G_\phi$. In a similar way, we find that there is an abrupt change appearing at a critical value of $G_\phi$ for sufficiently thin layers. The natural question is: what is the reason for these changes? An important clue is found in the fact that when the LDOS is fully suppressed, the odd-frequency correlations must be zero \cite{yokoyama_prb_07}. The presence of odd-frequency correlations will in general lead to an enhancement of the LDOS at zero-energy, which at present is one of the main suggestions put forth in the literature with regard to the issue of how to obtain clear experimental signatures of this exotic type of superconducting pairing. Therefore, the abrupt transition from a fully suppressed LDOS to a LDOS which is enhanced even compared to the normal-state value is a strong indicator of a symmetry-transition from the usual even-frequency correlations to a state of mixed even- and odd-frequency correlations, or possibly even pure odd-frequency correlations. It is therefore clear that the spin-DIPS occuring at the interface have paramount consequences with regard to the symmetry-properties of the induced superconducting correlations in the ferromagnet. Due to the complexity of the problem, it is unfortunately not possible to give an exact analytical treatment of the influence of $G_\phi$ on the symmetry-properties of the anomalous Green's function.

\begin{figure}[h!]
\centering
\resizebox{0.45\textwidth}{!}{
\includegraphics{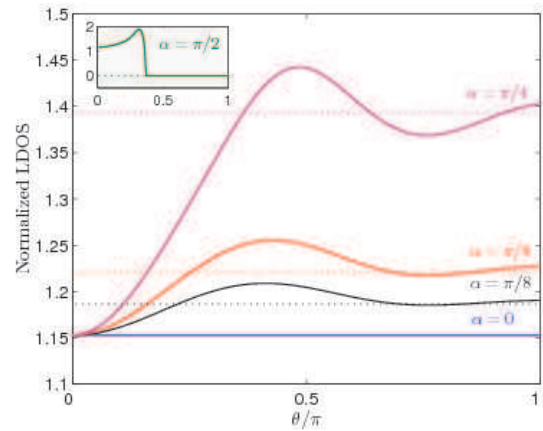}}
\caption{(Color online) Plot of the zero-energy LDOS at $x=d_F$ as a function of $\theta$ for several values of $\alpha$. We have fixed $h/\Delta_0=5, d_F/\xi_S=0.5, G_\phi=0$ to focus on the effect of the magnetic structure. The dotted lines give the result for $h_y=h_z=0$, which corresponds to the saturating behavior when $\theta$ increases since the average value of $h_y$ and $h_z$ vanishes in this limit. The inset shows the case $\alpha=\pi/2$, corresponding to $h_x=0$. For increasing $\theta$, the ferromagnetic layer effectively acts as a normal metal, thus causing a complete suppression of the zero-energy LDOS.}
\label{fig:conical_theta}
\end{figure}

In the remaining part of the discussion of conical ferromagnets, we wish to focus on how the proximity-induced LDOS depends on the structure of the magnetic texture, which is determined by the parameters $\{a,\alpha,\theta\}$ in Fig. \ref{fig:model}. We here focus on the role of $\alpha$ and $\theta$, which control respectively the direction and the speed of rotation of the magnetization upon entering the ferromagnetic layer. Thus, we keep $a/\xi_S$ fixed at $a/\xi_S=0.0263$. In Fig. \ref{fig:conical_theta}, we present results for the zero-energy LDOS at $x=d_F$ as a function of $\theta$ for several values of $\alpha$. The LDOS displays oscillations as a function of $\theta$, and eventually seems to sature upon increasing $\theta$. This may be understood microscopically by realizing that when the rotation of the magnetization texture becomes faster, \ie increasing $\theta$, the effective magnetization felt by the Cooper pair averages out to zero for the rotating components. For our setup, this would mean that only the $h_x$-component should remain non-zero, while $h_y=h_z=0$. To verify this scenario, we have also plotted the results in the $h_y=h_z=0$ case in Fig. \ref{fig:conical_theta} (dotted lines) for each value of $\alpha$, which is seen to coincide with the limiting behavior in the the high-$\theta$ case. It is interesting to note that for $\alpha=\pi/2$, the LDOS vanishes completely above a critical value for $\theta$. This may be understood by noting that $h_x=0$ when $\alpha=\pi/2$. Thus, when $\theta$ increases, we have $\langle h_x \rangle = \langle h_y \rangle = 0$, causing the ferromagnetic layer to act as a normal metal. 

\begin{figure}[b!]
\centering
\resizebox{0.5\textwidth}{!}{
\includegraphics{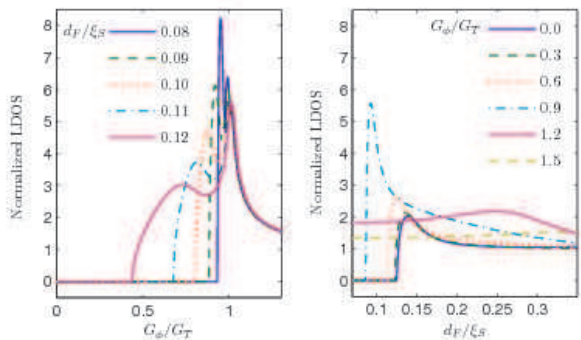}}
\caption{(Color online) Plot of the zero-energy LDOS at $x=d_F$ in the case of a Bloch domain wall with $h/\Delta_0=5$. In a) and c), we plot the LDOS as a function of the spin-DIPS $G_\phi$, while in b) and d) we plot it as a function of the ferromagnetic layer thickness $d_F$. For thin layers $d_F/\xi_S\ll1$, one observes an abrupt transition from a fully suppressed DOS to a non-zero DOS at a critical value for either $G_\phi$ or $d_F$.}
\label{fig:domainwall_crossover}
\end{figure}

\subsection{Discussion}\label{sec:discussionI}
The main approximation that we have made in our calculations is to use the bulk solution for the order parameter of the 
superconductor. Although this approximation is expected to be satisfactory in the regime $d_S\gg\{\xi_S,d_F\}$, such that 
the superconductor acts as a reservoir, there are two aspects which are lost upon doing so. One aspect is the depletion 
of the superconducting order parameter near the interface. The depletion may be disregarded in the tunneling limit 
\cite{bruder_prb_90} (low barrier transparency), and we do not expect that an inclusion of the spatial profile of 
the superconducting order parameter near the interface should have any qualitative influence upon our results, as 
long as the superconducting order parameter is not dramatically reduced at the interface. 
\par
\begin{figure}[b!]
\centering
\resizebox{0.45\textwidth}{!}{
\includegraphics{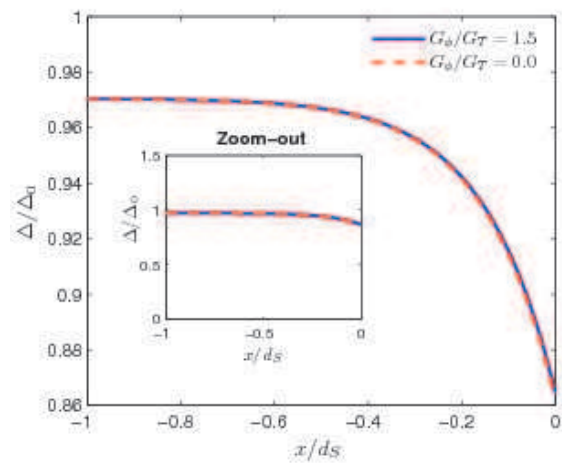}}
\caption{(Color online) Self-consistent solution for the spatial profile of the superconducting order parameter, using the approach described in Part II of this paper. The choice for parameter values are specified in the main text. To add stability to the numerical calculations, we added a small imaginary number $\i\eta$ to the quasiparticle energies $\varepsilon$, with $\eta=0.05\Delta_0$, effectively modeling inelastic scattering.}
\label{fig:gap}
\end{figure}

The assumption of a step-function superconducting order parameter is commonly employed in the literature, but let us for the sake of clarity here examine a bit more carefully under which circumstances this is truly warranted. In the present work, we have considered a superconducting reservoir of size $d_S\gg\xi_S$ and a ferromagnetic film of size $d_F \leq \xi_S$. For a weak ferromagnet considered here, the ferromagnetic coherence length $\xi_F$ is comparable in size to $\xi_S$. Also, we have considered the case where $\zeta=R_B/R_F\gg1$, corresponding to a low barrier transparency, which should be experimentally relevant. To investigate quantitatively how much the superconducting order parameter is suppressed near the interface, let us fix $h/\Delta_0=10$, $d_S/\xi_S=5$, $d_F/\xi_F=1$, and $\zeta=5$. Using a numerical approach for S$\mid$F bilayer with a homogeneous exchange field as employed in Part II of our paper, we obtain the gap self-consistently with the result shown in Fig. \ref{fig:gap}. It is also necessary to introduce the  barrier asymmetry factor $\gamma=\xi_S\sigma_F/(\xi_F\sigma_S)$, where $\sigma_{F(S)}$ is the conductivity in the F (S) layer. Here, we set $\gamma=1$. As seen, the depletion of the gap is quite insensitive to the value of $G_\phi$, and we have verified that the depletion of the gap is virtually the same even up to ferromagnetic layer thicknesses of $d_F/\xi_F=4$. As recently pointed out in Ref. \cite{cottet_arxiv_08}, the step-function approximation breaks down for low values of $\zeta$ and/or high values of $\gamma$, and if the spin-DIPS $G_\phi^S$ induced on the superconducting side are large in magnitude compared to the tunneling conductance $G_T$, the suppression of the gap becomes more pronounced.

The second aspect which is lost is the inverse proximity effect in the superconductor. The inverse proximity effect 
is, in similarity to the depletion of the order parameter, expected to be small when the interface transparency is 
low and $d_S \gg d_F$. Nevertheless, the presence of the spin-DIPS at the interface, modelled 
through the parameter $G_\phi$, could have some non-trivial impact on the correlations in the superconductor. Cottet showed that this 
may indeed be so in Ref.~\onlinecite{cottet_prb_07}, at least when the superconducting layer is quite thin. The full effect exerted on the LDOS by the presence of spin-DIPS on both sides of the interfaces was recently investigated numerically in an S$\mid$F bilayer \cite{cottet_arxiv_08}. However, no study so far have investigated how the proximity-induced magnetization in the superconducting region is affected by spin-DIPS. We will proceed to investigate this particular issue in detail in Part II of this work.
\par
Above, we have considered the diffusive limit $\xi_S/l_\text{imp}\gg1$, where $l_\text{imp}=v_F\tau$ is the mean 
free path. Although the magnetic texture we have considered in the second part is identical that of the conical 
ferromagnet Ho, one important difference is that Ho is a strong ferromagnet, contrary to the case studied here. This 
means that the diffusive limit condition $h\tau\ll1$ is not fulfilled for Ho, and it was in fact estimated in 
\cite{sosnin_prl_06} that $h\tau \simeq 10$. This calls for a treatment with the more general Eilenberger equation, 
which allows for a study where the energy scale of the Zeeman-splitting is comparable or larger than the 
self-energy associated with impurity scattering. A natural continuation of this work would therefore be to study a 
proximity-structure of a superconductor$\mid$conical ferromagnet for an arbitrary ratio of the parameter $h\tau$. 
Such an endeavor would nevertheless be quite challenging unless a weak proximity effect is assumed. In the present 
work, we have not restricted ourselves to any limits with regard to the barrier transparency or the proximity 
effect. Although the exchange field considered for the conical ferromagnet in this paper is smaller than the one 
realized in Ho, we expect that our results may be qualitatively relevant for STM-measurements in superconducting junctions with Ho. In general, 
increasing the exchange field amounts to a quantitative reduction of the magnitude of the proximity effect. 
\par
Finally, we show that the zero-energy DOS for the domain wall case exhibits a similar crossover behavior as the 
conical ferromagnetic case upon varying $G_\phi$ and $d_F$ when $d_F/\xi_S \ll 1$. In 
Fig. \ref{fig:domainwall_crossover}, the zero-energy DOS is plotted for the thick-domain wall case to 
illustrate this effect - the results are very similar even for $d_W/d_F\ll1$ when $d_F/\xi_S\ll1$. 
Once again, it should be noted that a complete suppression of the DOS amounts to pure even-frequency 
superconducting correlations induced in the ferromagnetic region, since the presence of odd-frequency 
correlations enhances the zero-energy DOS. The exact microscopic mechanism behind the abrupt crossover 
occuring at critical values of $G_\phi$ and $d_F$, respectively, remains somewhat unclear. A possible 
resolution to this behavior is the observation that the spin-DIPS may conspire with the proximity-induced 
minigap in the ferromagnetic region for sufficiently thin layers ($d_F/\xi_S\ll1$) and yield a zero-energy 
DOS of the form $N(0) \sim 1/\sqrt{G_\phi^2-G_T^2}$, as noted in Ref. \cite{huertashernando_prl_02}. In 
this case, a scenario similar to the one of a thin-film superconductor in the presence of an in-plane 
magnetic field is realized, where the spin-resolved DOS experiences a quasiparticle energy-shift 
with $\pm h$. In this case, the role of the exchange field is played by $G_\phi$ while the role of 
the superconducting gap is played by $G_T$. We do not observe the effects shown in 
Fig. \ref{fig:domainwall_crossover} for larger values of $d_F$, which is consistent with the fact 
that the minigap is completely absent in this regime since the proximity effect becomes weaker.

\section{Inverse proximity effect in a S$\mid$F bilayer with a homogeneous magnetization texture}

\begin{figure}[b!]
\centering
\resizebox{0.5\textwidth}{!}{
\includegraphics{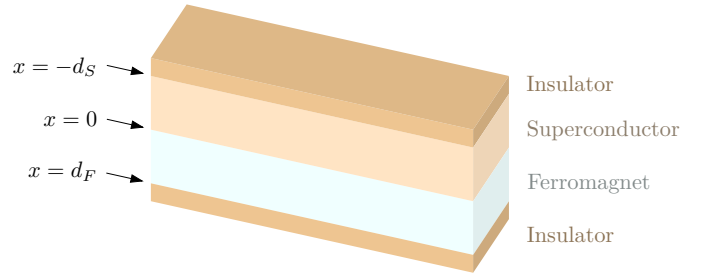}}
\caption{(Color online) The experimental setup proposed in this paper: a superconductor$\mid$ferromagnet bilayer.}
\label{fig:model2}
\end{figure}

In this part of the paper, we will consider the inverse proximity effect of an S$\mid$F bilayer, where the exchange field is fixed and parallel to the $z$-axis, manifested through an induced magnetization near the interface of the superconducting region. We will again employ the quasiclassical theory of superconductivity \cite{serene}, and consider the diffusive limit described by the Usadel equation \cite{usadel}, as this is experimentally the most relevant case. Our approach will be to solve the Usadel equation and the gap equation for the superconducting order parameter self-consistently everywhere in the system shown in Fig. \ref{fig:model2}. 

\subsection{Theory}\label{sec:theoryII}

We will use the conventions and notation of Ref. \cite{linder_prb_08_2}, which also allows for an inclusion of magnetic impurities and spin-orbit coupling if desirable. To facilitate the numerical implementation, we employ the following parametrization of the Green's functions:
\begin{align}\label{eq:green}
\hat{g}_j = \begin{pmatrix}
c_{\uparrow,j} & 0 & 0 & s_{\uparrow,j}\\
0 & c_{\downarrow,j} & s_{\downarrow,j} & 0 \\
0 & -s_{\downarrow,j} & -c_{\downarrow,j} & 0 \\
-s_{\uparrow,j} & 0 & 0 & -c_{\uparrow,j} \\
\end{pmatrix},\; j=\{S,F\}
\end{align}
where we have introduced 
\begin{align}
s_{\sigma,j} = \sinh(\theta_{\sigma,j}),\; c_{\sigma,j} = \cosh(\theta_{\sigma,j}).
\end{align}
Note that $(\hat{g}_j)^2=\hat{1}$ is satisfied. The parameter $\theta_{\sigma,j}$ is a measure of the proximity effect, and obeys the Usadel equation
\begin{align}\label{eq:usadel}
D_j\partial_x^2\theta_{\sigma,j} &+ 2\i(\varepsilon+\sigma h)\sinh(\theta_{\sigma,j})  \notag\\
&- 2\i\sigma\Delta\cosh(\theta_{\sigma,j}) = 0,\;  \sigma=\{\uparrow,\downarrow\}
\end{align}
in the superconducting ($h=0, j=S$) and ferromagnetic ($\Delta=0, j=F$) layer, respectively. Above, $D_S$ and $D_F$ denote the diffusion constants in the two layers, $\varepsilon$ is the quasiparticle energy, $\Delta$ is the pair potential, while $h$ is the exchange field. The two latter are in general subject to a depletion close to the S$\mid$F interface.
\par
The boundary condition for the ferromagnetic Green's function, $\hat{g}_F$, reads \cite{huertashernando_prl_02}
\begin{align}\label{eq:bcF}
2\xi_F\hat{g}_F \partial_x \hat{g}_F = \gamma_T[\hat{g}_S, \hat{g}_F] + \i \gamma_{\phi,F} [\hat{\alpha}_3, \hat{g}_F]
\end{align} 
at $x=0$, and $ \hat{g}_F\partial_x\hat{g}_F=\hat{0}$ at $x=d_F$. Here, $\hat{\ldots}$ denotes a $4\times4$ matrix in spin$\otimes$particle-hole space. Also, $\hat{\alpha}_3=\text{diag}(1,-1,1,-1)$. For the superconducting Green's function, $\hat{g}_S$, we have
\begin{align}\label{eq:bcS}
2(\xi_S/\gamma)\hat{g}_S \partial_x \hat{g}_S = -\gamma_T[\hat{g}_F, \hat{g}_S] - \i \gamma_{\phi,S} [\hat{\alpha}_3, \hat{g}_S]
\end{align} 
at $x=0$, and $ \hat{g}_S\partial_x\hat{g}_S=\hat{0}$ at $x=-d_S$. Above, we have defined 
\begin{align}
\gamma_T = G_T\xi_F/(A\sigma_F),\; \gamma_{\phi,F(S)} = G_{\phi,F(S)}\xi_F/(A\sigma_F),
\end{align}
and the barrier asymmetry factor 
\begin{align}
\gamma = \xi_S\sigma_F/(\xi_F\sigma_S).
\end{align}
Moreover, $A$ is the tunneling contact area, while $\sigma_{F(S)}$ are the normal-state conductivities. Note that 
\begin{align}
A\sigma_{F(S)} = d_{F(S)}/R_{F(S)},
\end{align}
where $d_{F(S)}$ is the thickness of the layer and $R_{F(S)}$ is the normal-state resistance. 
\par
In total, the interface between the S and F regions is thus characterized by three parameters: the normalized barrier conductance $\gamma_T$, the spin-DIPS $\gamma_{\phi,S}$ and $\gamma_{\phi,F}$ on each side of the interface. 
In what follows, we will study the mutual influence of superconductivity and ferromagnetism on each other, instead of assuming the bulk solution for $\hat{g}_S$ in the superconducting region, as is usually done in the literature. We solve the Usadel equation self-consistently in both the S and F layer, supplementing it with the gap equation:
\begin{align}
\Delta = \frac{N_F\lambda}{2}\int^\omega_0\text{d}\varepsilon \tanh{(\beta\varepsilon/2)} \sum_\sigma\sigma \text{Re}\{\sinh(\theta_\sigma)\},
\end{align}
where we choose the weak coupling-constant and cut-off energy to be $N_F\lambda=0.2$ and $\omega/\Delta_0 = 75$.
When obtaining the Green's functions, a number of interesting physical quantities may be calculated. For instance, the normalized LDOS is obtained according to 
\begin{align}
N(\varepsilon)/N_0 = \text{Re}\{\cosh\theta_\uparrow + \cosh\theta_\downarrow\}/2.
\end{align}
Experimentally, the LDOS may be probed at $x=-d_S$ in the superconducting layer and $x=d_F$ in the ferromagnetic layer by performing tunneling spectroscopy through the insulating layer. In principle, it is also possible to obtain the LDOS at any position $x$ by using spatially-resolved scanning tunneling microscopy. 
\par
The quantity of interest which we shall focus on in this work is the proximity-induced magnetization in the superconducting region. A few words about the sign of the magnetization in the problem is appropriate. First, recall that the magnetic moment $\boldsymbol{\mu}$ of an electron is directed \textit{opposite} to its spin $\mathbf{S}$, namely $\boldsymbol{\mu} \simeq -(e/m_e)\boldsymbol{S}$, where $e=|e|$ and $m_e$ is the electron charge and mass. Therefore, if the exchange energy $h$ favors spin-$\uparrow$ electrons energetically, the resulting magnetization $\boldsymbol{M}$ of the ferromagnet will be directed in the opposite direction, $\boldsymbol{M}\parallel (-\boldsymbol{z})$. 
\par
In the absence of a proximity effect, we have $\boldsymbol{M}=0$ in the superconducting region and $\boldsymbol{M} = M_0\hat{\boldsymbol{z}}$ in the ferromagnetic region, where 
\begin{align}
M_0 \simeq -\mu_BN_0h
\end{align}
in the quasiclassical approximation $h\ll \varepsilon_F$. Now, the change in magnetization due to the proximity effect may be calculated according to
 \begin{align}
\delta\boldsymbol{M} = -\mu_B \hat{\boldsymbol{z}}\sum_\sigma \sigma \langle \psi_\sigma^\dag \psi_\sigma \rangle
\end{align}
 in both the superconducting and ferromagnetic region. Using a quasiclassical approach, the above expression translates into a normalized change in magnetization
\begin{align}\label{eq:mag}
\delta M/M_0 = -\int^\infty_0 \frac{\text{d}\varepsilon}{h} \sum_\sigma \sigma \text{Re} \{\cosh\theta_\sigma\}\tanh(\beta \varepsilon/2).
\end{align}
In the ferromagnetic region, the normalized magnetization $M/M_0$ is therefore $1 + \delta M_F/M_0$, while in the superconducting region we have an induced magnetization $\delta M_S/M_0$, where $\delta M_{F(S)}$ is determined by Eq. (\ref{eq:mag}) on the ferromagnetic (superconducting) side of the interface. 

\begin{figure}[t!]
\centering
\resizebox{0.5\textwidth}{!}{
\includegraphics{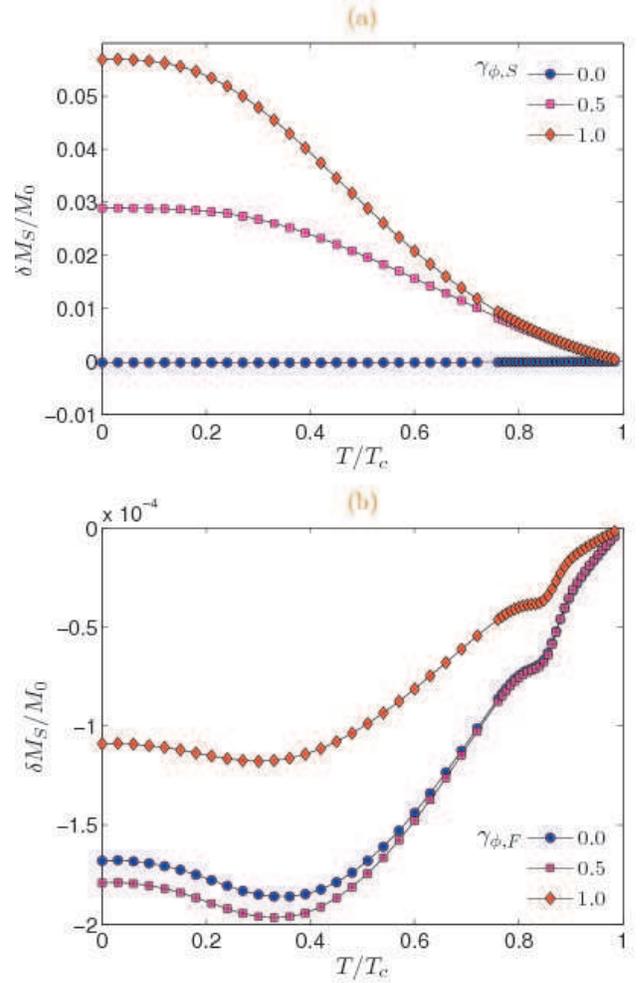}}
\caption{(Color online) (a) Plot of the proximity-induced magnetization for $\gamma_{\phi,F}=0$ upon varying the spin-DIPS $\gamma_{\phi,S}$ on the superconducting side. (b) Plot of the proximity-induced magnetization for $\gamma_{\phi,S}=0$ upon varying the spin-DIPS $\gamma_{\phi,F}$ on the ferromagnetic side. We have used $d_S/\xi_S=0.2$, and the other parameter values are discussed and provided in the main text. Note that the lines with a (blue) circle marker are equal in (a) and (b), corresponding to $\gamma_{\phi,S}=\gamma_{\phi,F}=0$. As seen, $\gamma_{\phi,S}$ affects $\delta M_S/M_0$ much more strongly than $\gamma_{\phi,F}$. In (b), the magnetization switches sign upon increasing $\gamma_{\phi,F}$ further (for the present parameters, the sign switch occurs around $\gamma_{\phi,F} \simeq 1.3$).}
\label{fig:Mag_T}
\end{figure}
\par

Although we shall be concerned with a full numerical solution when presenting our results in Sec. \ref{sec:resultsII}, let us for completeness sketch how an analytical solution may be obtained under the assumption of a weak proximity effect. Including the spin-DIPS, the analytical results obtained here are thus a natural extension of the results in Ref. \cite{bergeret_prb_04}, where the spin-DIPS were neglected. We remind the reader that spin-DIPS occur whenever there is a finite spin-polarization in the ferromagnetic region or when the barrier itself is magnetic. 
\par
In the weak-proximity regime, the Usadel equation in the ferromagnetic region becomes
\begin{align}
D_F\partial_x^2\delta\theta_\sigma^F + 2\i(\varepsilon+\sigma h)\delta\theta_\sigma^F = 0,
\end{align}
where the linearization of Eqs. (\ref{eq:green}) and (\ref{eq:usadel}) amounts to $\theta_{\sigma,F}\to \delta\theta_\sigma^F$ where $|\delta\theta_\sigma^F|\ll1$. The general solution is readily obtained as
\begin{align}\label{eq:weakF}
\delta\theta_\sigma^F = A_\sigma(\e{\i k_\sigma x} + \e{-\i k_\sigma x + 2\i k_\sigma d_F}),
\end{align}
upon taking into account the vacuum boundary condition $\partial_x \delta\theta_\sigma^F=0$ at $x=d_F$, and defining
\begin{align}
k_\sigma^2 &= 2\i(\varepsilon+\sigma h)/D_F.
\end{align}
 In the superconducting region, we obtain the Usadel equation
\begin{align}
D_S\partial_x^2\delta\theta_\sigma^S + 2\i(\varepsilon c_\text{BCS} - \Delta s_\text{BCS}) = 0,
\end{align}
under the assumption that the superconducting order parameter is virtually unaltered from the bulk case. This is a valid approximation for $\{\gamma,\gamma_T\}\ll1$ and not too large $\gamma_{\phi,S}$ (typically $\gamma_{\phi,S}<1$), which we have verified by using the full numerical solution. Here, $\delta\theta_\sigma^S$ is the deviation from the bulk BCS solution, i.e. $\theta_{\sigma,S} \to \sigma\theta_\text{BCS} + \delta\theta_\sigma^S$ with $|\delta\theta_\sigma^S|\ll1$ and 
\begin{align}
c_\text{BCS} = &\cosh(\theta_\text{BCS}),\; s_\text{BCS} = \sinh(\theta_\text{BCS}),\notag\\
 &\theta_\text{BCS} = \text{atanh}(\Delta/\varepsilon).
\end{align}
In this case, the general solution reads
\begin{align}\label{eq:weakS}
\delta\theta_\sigma^S = B_\sigma(\e{\i \kappa x} + \e{-\i\kappa x - 2\i \kappa d_S}),
\end{align}
when incorporating the vacuum boundary condition $\partial_x \delta\theta_\sigma^S=0$ at $x=-d_S$, upon defining
\begin{align}
\kappa^2 = (\varepsilon c_\text{BCS} - \Delta s_\text{BCS})/D_S.
\end{align}
The remaining task is to determine the unknown coefficients $\{A_\sigma,B_\sigma\}$. Linearizing the boundary conditions Eq. (\ref{eq:bcF}) and (\ref{eq:bcS}), we obtain at $x=0$
\begin{align}
\xi_S \partial_x\delta\theta_\sigma^S/\gamma &= \gamma_T(c\delta\theta_\sigma^F -\sigma s_\text{BCS} - c_\text{BCS}\delta\theta_\sigma^S) \notag\\
&- \sigma\i\gamma_{\phi,S}(\sigma s_\text{BCS} + c_\text{BCS}\delta\theta_\sigma^S),\notag\\
\xi_F\partial_x\delta\theta_\sigma^F &= \gamma_T(c\delta\theta_\sigma^F -\sigma s_\text{BCS} - c_\text{BCS}\delta\theta_\sigma^S)\notag\\
&+\sigma\i\gamma_{\phi,F}\delta\theta_\sigma^F.
\end{align}
From these boundary conditions, one derives that
\begin{align}\label{eq:coeff}
A_\sigma &= \frac{z_4^\sigma}{z_3^\sigma}\frac{s_\text{BCS}[z_3^\sigma(\gamma_T\sigma + \i\gamma_{\phi,S}) - z_1^\sigma\sigma\gamma_T]}{z_2^\sigma z_3^\sigma - z_1^\sigma z_4^\sigma} - \sigma s_\text{BCS} \gamma_T,\notag\\
B_\sigma &= \frac{s_\text{BCS} [z_1^\sigma \sigma\gamma_T - z_3^\sigma(\gamma_T\sigma + \i\gamma_{\phi,S})]}{z_2^\sigma z_3^\sigma - z_1^\sigma z_4^\sigma}.
\end{align}
Here, we have defined the auxiliary quantities:
\begin{align}
z_1^\sigma &= -\gamma_Tc_\text{BCS}(1+\e{2\i k_\sigma d_F})\notag\\
z_2^\sigma &= \frac{\i \kappa \xi_S(1-\e{-2\i \kappa d_S})}{\gamma} + c_\text{BCS}(\gamma_T + \i\sigma \gamma_{\phi,S})(1+\e{-2\i\kappa d_S})\notag\\
z_3^\sigma &= \i k_\sigma \xi_F(1-\e{2\i k_\sigma d_F}) - (\gamma_Tc_\text{BCS} + \sigma \i\gamma_{\phi,F})(1+\e{2\i k_\sigma d_F})\notag\\
z_4^\sigma &= c_\text{BCS}\gamma_T(1+\e{-2\i \kappa d_S}).
\end{align}
Eqs. (\ref{eq:weakF}), (\ref{eq:weakS}), and (\ref{eq:coeff}) constitute a closed analytical solution for the Green's functions in the entire S$\mid$F bilayer. To use this analytical solution, one should verify that $|\delta\theta_\sigma^{F,S}|\ll1$ for the relevant parameter regime. Spin-flip and spin-orbit scattering may also be accounted for in the analytical solution of the Green's function by adding appropriate terms to the Usadel equation. The calculation is then performed along the lines of Refs. \cite{linder_prb_08_2, linder_spin}.

\begin{figure}[t!]
\centering
\resizebox{0.5\textwidth}{!}{
\includegraphics{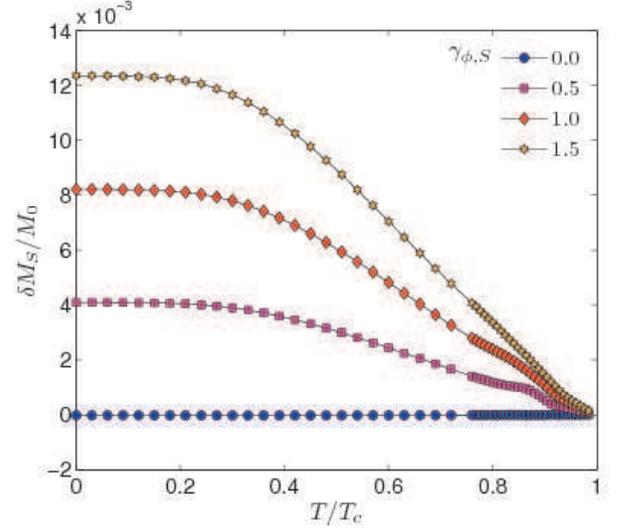}}
\caption{(Color online) Plot of the proximity-induced magnetization for $\gamma_{\phi,F}=0$ upon varying the spin-DIPS $\gamma_{\phi,S}$ on the superconducting side using $d_S/\xi_S=1.0$. }
\label{fig:combination}
\end{figure}

\begin{figure}[t!]
\centering
\resizebox{0.5\textwidth}{!}{
\includegraphics{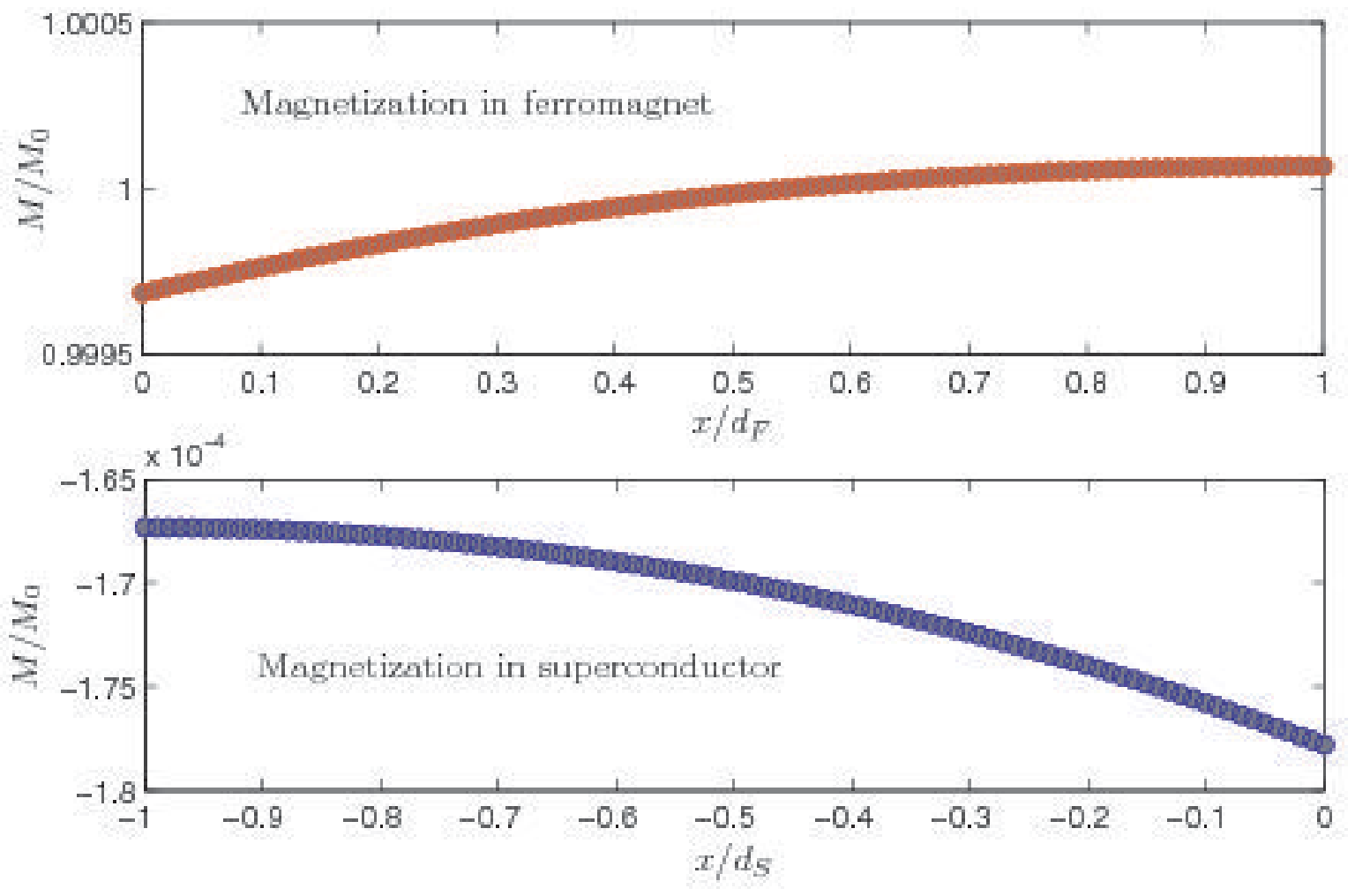}}
\caption{(Color online) Plot of the spatial dependence of the total magnetization at zero temperature for $d_S/\xi_S=0.2$ and $\gamma_{\phi,F}=\gamma_{\phi,S}=0.0$. }
\label{fig:spatial0.0}
\end{figure}

\begin{figure}[t!]
\centering
\resizebox{0.5\textwidth}{!}{
\includegraphics{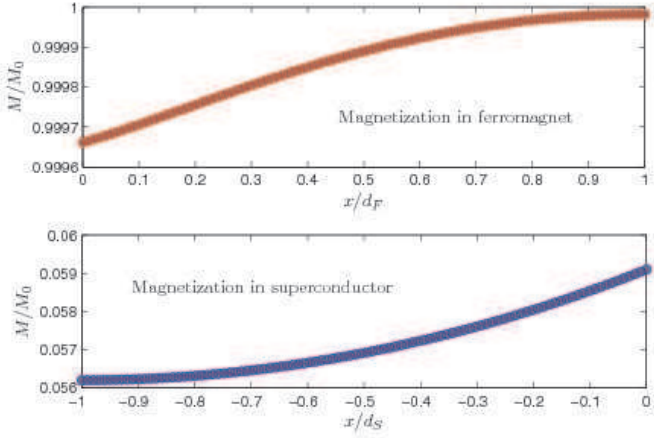}}
\caption{(Color online) Plot of the spatial dependence of the total magnetization at zero temperature for $d_S/\xi_S=0.2$, $\gamma_{\phi,F}=0.0$ and $\gamma_{\phi,S}=1.0$. }
\label{fig:spatial1.0}
\end{figure}

\subsection{Results}\label{sec:resultsII}
We are now in a position to evaluate the proximity-induced magnetization numerically. The full (non-linearized) Usadel equation will be employed, such that we are not restricted to the weak-proximity effect regime. To stabilize the numerical calculations, we add a small imaginary number to the quasiparticle energy, $\varepsilon\to \varepsilon +\i\eta$, with  $\eta=0.05\Delta_0$. We focus on the results reported very recently by Xia \etal, \cite{xia_arxiv_08} and take $d_S/\xi_S=0.2$, $d_F/\xi_F=1.0$, and $h/\Delta_0=15$ as a reasonable set of parameters which should be relevant to this experiment. Also, we assume that the junction conductance was low, $\gamma_T=0.1$, and we set the barrier asymmetry factor to $\gamma=0.2$, corresponding to a scenario where the superconducting region is much less disordered than the ferromagnetic one. We will also investigate the case $d_S/\xi_S=1.0$, to see how the properties of the system changes when going away from the limit $d_S/\xi_S\ll 1$. We underline that our main objective in this work is to investigate the influence of the spin-DIPS on the proximity-induced magnetization in the system, such that we mainly vary $\gamma_{\phi,F}$ and $\gamma_{\phi,S}$ while keeping the other parameters fixed.
\par
Let us first consider the temperature-dependence of the proximity-induced magnetization in the superconducting region in Fig. \ref{fig:Mag_T}. To clarify the role of the spin-DIPS on each side of the interface, we plot $\delta M_S/M_0$ for several values of $\gamma_{\phi,S}$ in Fig. \ref{fig:Mag_T} (a) while keeping $\gamma_{\phi,F}=0$ fixed. Conversely, we plot $\delta M_S/M_0$ for several $\gamma_{\phi,F}$ in Fig. \ref{fig:Mag_T} (b) with $\gamma_{\phi,S}=0$. In both cases, we plot the proximity-induced magnetization at $x=-d_S$. One obvious difference between these two scenarios is that the spin-DIPS on the superconducting side, $\gamma_{\phi,S}$, influence the proximity-induced magnetization much stronger than $\gamma_{\phi,F}$. The same thing is true with regard to the influence of spin-DIPS on the superconducting order parameter: $\gamma_{\phi,S}$ influences the spatial profile of $\Delta$ much more than what $\gamma_{\phi,F}$ does. From Fig. \ref{fig:Mag_T}, it is clearly seen how the proximity-induced magnetization may switch sign upon increasing the magnitude of the spin-DIPS $\gamma_{\phi,S}$. We have checked numerically that this effect also takes upon increasing $\gamma_{\phi,F}$ when keeping $\gamma_{\phi,S}=0$. \textit{Thus, increasing either} $\gamma_{\phi,S}$ \textit{or} $\gamma_{\phi,F}$ \textit{can lead to a sign change of the proximity-induced magnetization in the superconducting region.} It is then clear that the conclusion of Ref. \cite{kharitonov_prb_06} that only spin screening is possible in diffusive S$\mid$F bilayers does not hold in general, since the presence of spin-DIPS alters the screening effect. In what follows, we focus on the role of $\gamma_{\phi,S}$ since its impact on $\delta M_S/M_0$ is much greater than that of $\gamma_{\phi,F}$. In Fig. \ref{fig:combination}, we consider the case $d_S/\xi_S=1.0$ to show that the sign change of the magnetization persists when going away from the limit $d_S/\xi_S\ll1$. The spatial profile of the total magnetization in the F and S regions are shown for the case $d_S/\xi=0.2$ with $\gamma_{\phi,S}=0.0$ in Fig. \ref{fig:spatial0.0} and $\gamma_{\phi,S}=1.0$ in Fig. \ref{fig:spatial1.0}. It is seen that the magnetization decreases in a monotonic fashion toward the superconducting region, and reaches its bulk value deep inside the ferromagnetic region. In the superconductor, magnetization is induced near the interface and decays  with the distance from the interface.

\begin{figure}[b!]
\centering
\resizebox{0.5\textwidth}{!}{
\includegraphics{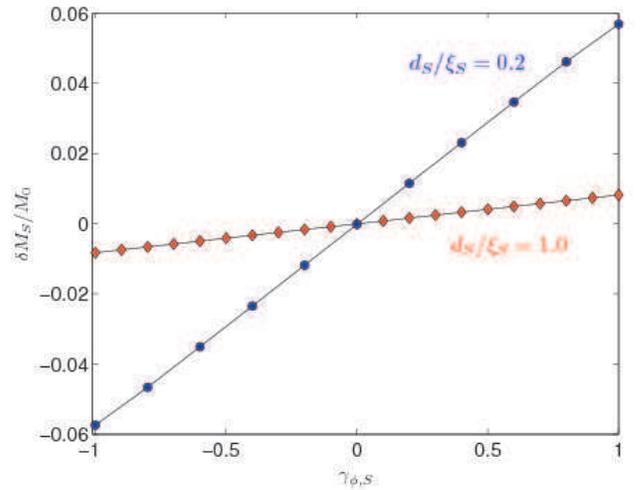}}
\caption{(Color online) Plot of the proximity-induced magnetization at $T=0$ as a function of $\gamma_{\phi,S}$ for two different values of $d_S/\xi_S$. }
\label{fig:gammaphi}
\end{figure}

\subsection{Discussion}\label{sec:discussionII}

We propose the following explanation for the anti-screening effect observed upon increasing $\gamma_{\phi,S}$. The effect of the spin-DIPS in 
the case of a thin superconducting layer $d_S\ll\xi_S$ in Ref. \cite{cottet_prb_07}  was shown to be equivalent to an internal magnetic 
exchange splitting $h_\text{eff}$ in the superconducting region. Therefore, the magnitude of the magnetization in the superconductor 
should essentially grow with an increasing value of $\gamma_{\phi,S}$. If this is the case, the proximity-induced magnetization should 
also be sensitive to \textit{the sign} of $\gamma_{\phi,S}$, as the opposite spin species would be energetically favored when comparing 
the case $\gamma_{\phi,S}$ with $(-\gamma_{\phi,S})$. 
To test this hypothesis, we plot in Fig. \ref{fig:gammaphi} the proximity-induced magnetization at $T=0$ as a function of the spin-DIPS 
on the superconducting side, $\gamma_{\phi,S}$ (keeping $\gamma_{\phi,F}=0$). The results confirm our hypothesis -- it is seen that 
$\delta M_S/M_0$ is an antisymmetric function of $\gamma_{\phi,S}$. The influence of $\gamma_{\phi,S}$ can also be seen directly in 
the LDOS in the superconducting region. For $\gamma_{\phi,S}\neq0$, we obtain a double-peak structure in the LDOS at $x=-d_S$ in 
agreement with Refs. \cite{cottet_prb_07, cottet_arxiv_08}, while the superconducting order parameter depletes very little close 
to the interface for the chosen parameter values. In general, the depletion of the superconducting order parameter is found to be 
small as long as $\{\gamma_T,\gamma\}\ll1$ and $\gamma_{\phi,S} \simeq 1$ or smaller.
\par
In Ref. \cite{bergeret_prb_04}, the inverse proximity effect of an S$\mid$F bilayer was studied without taking into account the 
presence of spin-DIPS, with the result that the proximity-induced magnetization in the superconducting region would have the opposite 
sign of the proximity ferromagnet, i.e. a screening effect. It was proposed in Ref. \cite{bergeret_prb_04} that this behavior could 
be understood physically by considering the contribution to the magnetization from Cooper pairs which were close to the interface: 
the spin-$\uparrow$ electron would prefer to be in the ferromagnetic region due to the exchange energy, while the spin-$\downarrow$ 
electron remaining in the superconducting region then would give rise to a magnetization in the opposite direction of the proximity 
ferromagnet. However, it is clear from the present study that this simple picture must be modified when properly considering the 
spin-DIPS $\gamma_{\phi,S}$ on the superconducting side of the junction, since they act as an effective exchange field 
inside the superconductor.
\par
In this paper, we have evaluated the proximity-induced magnetization in the vicinity of the interface without taking into account 
the Meissner response of the superconductor. This should be permissable in a thin-film geometry as the one employed in 
Ref. \cite{xia_arxiv_08}, where the screening currents are suppressed. In particular, for a field in the plane of the 
superconducting film (see Fig. \ref{fig:model2}), the Meissner effect should be strongly suppressed \cite{meservey} for 
$d_S/\xi_S\ll1$.

\section{Summary}\label{sec:summary}
In conclusion, we have in Part I of this work investigated the proximity effect in a superconductor$\mid$inhomogeneous ferromagnet junctions. 
Proper boundary conditions which take into account the spin dependent phase-shifts experienced by the reflected and 
transmitted quasiparticles were employed. As an application of our model, we have studied the LDOS in the ferromagnet in the presence of 
domain walls and a conical magnetic structure. We find that the presence of a domain wall enhances the odd-frequency correlations 
induced in the ferromagnet, manifested through a zero-energy peak in the LDOS. For the conical ferromagnet, we 
show that the spin-dependent  phase shifts originating with the interface have a strong qualitative effect on the LDOS, especially for thin layers. 
In particular, we find an abrupt crossover from a fully suppressed LDOS to a finite LDOS which appears at a critical thickness of 
the ferromagnetic layer, and the particular value of the critical thickness depends on the value of $G_\phi$. In a similar way, we 
find that there is an abrupt change appearing at a critical value of $G_\phi$ for sufficiently thin layers. We speculate that the 
reason for this could be a symmetry-transition from even- to odd-frequency correlations for the proximity-amplitudes in the 
ferromagnetic region. The theory developed in the present paper takes into account both the phase-shifts acquired by scattered 
quasiparticles at the 
interface due to the presence of ferromagnetic correlations, and also an arbitrary inhomogeneity of the magnetic texture on the 
ferromagnetic side. Our results for the conical ferromagnetic structure should be relevant for the material Ho, which was 
used in Ref. \cite{sosnin_prl_06} to indicate the presence of long-range superconducting correlations. 
\par
In Part II of this work, we have investiged numerically and self-consistently the inverse proximity effect in 
a superconductor$\mid$ferromagnet (S$\mid$F) bilayer, manifested through an induced magnetization in the superconducting 
region. We find that the interface properties play a crucial role in this context, as the spin-dependent interfacial 
phase-shifts (spin-DIPS) may invert the sign of the proximity-induced magnetization. This finding modifies previous 
conclusions obtained in the literature, and suggests that the influence of the spin-DIPS should be properly accounted 
for in a theory for the inverse proximity effect in S$\mid$F bilayers.

\acknowledgments
J.L. acknowledges M. Eschrig and A. Cottet for very useful discussions. J.L. and A.S. were supported by the Research Council 
of Norway, Grants No. 158518/432 and No. 158547/431 (NANOMAT), and Grant No. 167498/V30 (STORFORSK). T.Y. acknowledges 
support by JSPS.

\appendix

\section{Spin-active boundary conditions}
\begin{figure}[b!]
\centering
\resizebox{0.4\textwidth}{!}{
\includegraphics{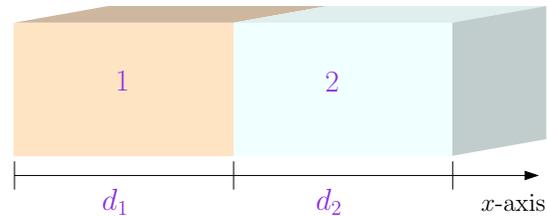}}
\caption{(Color online) Junction consisting of two regions 1 and 2 with an interface perpendicular to the $\boldsymbol{x}$-axis. }
\label{fig:appendix}
\end{figure}

To facilitate and encourage use of the spin-active boundary conditions required for an S$\mid$F interface, we here write down their 
explicit form in the diffusive limit for the case of a magnetization in the $\boldsymbol{z}$-direction, following 
Ref. \cite{huertashernando_prl_02, cottet_prb_07}. Consider a junction consisting of two regions 1 and 2, as shown 
in Fig. \ref{fig:appendix}. The regions have widths $d_j$ and bulk electrical resistances $R_j$. The matrices used 
below are $4\times4$ matrices in particle-hole$\otimes$spin space, using a basis 
\begin{align}
\psi(\vecr,t) = \begin{pmatrix}
\psi_\uparrow(\vecr,t)\notag\\
\psi_\downarrow(\vecr,t)\notag\\
\psi_\uparrow^\dag(\vecr,t)\notag\\
\psi_\downarrow^\dag(\vecr,t)\notag\\
\end{pmatrix}.
\end{align}
Introducing $\hat{\alpha} = \text{diag}(1,-1,1,-1) = \text{diag}(\underline{\sigma}_3,\underline{\sigma}_3)$, where $\underline{\sigma_3}$ is the third Pauli matrix in spin-space, we may write the boundary conditions as follows:
\begin{align}\label{eq:bcspin}
2(d_1/R_1) \hat{g}_1\partial_x \hat{g}_1 = G_T[\hat{g}_1,\hat{g}_2] - \i G_{\phi,1}[\hat{\alpha},\hat{g}_1],\notag\\
2(d_2/R_2) \hat{g}_2 \partial_x \hat{g}_2 = G_T[\hat{g}_1,\hat{g}_2] + \i G_{\phi,2}[\hat{\alpha},\hat{g}_2].
\end{align}
Here, $G_T$ is the conductance of the junction while $G_{\phi,j}$ are the phase-shifts on side $j$ of the interface. The 
parameters $\{G_T,G_{\phi,j}\}$ may be calculated by relating them to microscopic transmission and reflection probabilities 
within \eg a Blonder-Tinkham-Klapwijk (BTK) \cite{btk} framework. Explicitly spin-active barriers were considered in 
ballistic S$\mid$F bilayers using the BTK-approach for both $s$-wave \cite{linder_prb_07} and $d$-wave 
\cite{kashiwaya_prb_99} superconductors. In the absence of spin-DIPS $(G_{\phi,j}\to0)$, Eq. (\ref{eq:bcspin}) reduce 
to the Kupriyanov-Lukichev non-magnetic boundary conditions \cite{kupluk}. Let us make a final remark concerning the 
treatment of interfaces in the quasiclassical theory of superconductivity. We previously stated that the application of 
the present theory requires that the characteristic energies of various self-energies and perturbations in the system 
are much smaller than the Fermi energy $\varepsilon_\text{F}$. At first glance, this might seem to be irreconcilable 
with the presence of interfaces, which represent strong perturbations varying on atomic length scales. However, this 
problem may be overcome by including the interfaces as boundary conditions for the Green's functions rather than 
directly as self-energies in the Usadel equation.

\end{document}